\documentclass[aps,pre,twocolumn,superscriptaddress,showpacs]{revtex4}
\usepackage{graphics}

\begin{document}

\title{Vaccination against rubella: Analysis of the temporal evolution of the age-dependent 
force of infection and the effects of different contact patterns}

\author{M. Amaku}
\thanks{Author for correspondence. E-mail: amaku@vps.fmvz.usp.br} 
\affiliation{Departamento de Medicina Veterin\'{a}ria Preventiva 
e Sa\'{u}de Animal,Faculdade de Medicina Veterin\'{a}ria e Zootecnia, 
Universidade de S\~{a}o Paulo, Av. Prof. Dr. Orlando Marques de Paiva 87, 
S\~{a}o Paulo, SP, 05508-000, Brazil}
\affiliation{Faculdade de Medicina, Universidade de S\~{a}o Paulo,
and LIM01/HCFMUSP, Av. Dr. Arnaldo 455, S\~{a}o Paulo, SP, 01246-903, Brazil}
\author{F. A. B. Coutinho}
\affiliation{Faculdade de Medicina, Universidade de S\~{a}o Paulo,
and LIM01/HCFMUSP, Av. Dr. Arnaldo 455, S\~{a}o Paulo, SP, 01246-903, Brazil}
\author{R. S. Azevedo}
\affiliation{Faculdade de Medicina, Universidade de S\~{a}o Paulo,
and LIM01/HCFMUSP, Av. Dr. Arnaldo 455, S\~{a}o Paulo, SP, 01246-903, Brazil}
\author{M. N. Burattini}
\affiliation{Faculdade de Medicina, Universidade de S\~{a}o Paulo,
and LIM01/HCFMUSP, Av. Dr. Arnaldo 455, S\~{a}o Paulo, SP, 01246-903, Brazil}
\author{L. F. Lopez}
\affiliation{Faculdade de Medicina, Universidade de S\~{a}o Paulo,
and LIM01/HCFMUSP, Av. Dr. Arnaldo 455, S\~{a}o Paulo, SP, 01246-903, Brazil}
\author{E. Massad}
\affiliation{Faculdade de Medicina, Universidade de S\~{a}o Paulo,
and LIM01/HCFMUSP, Av. Dr. Arnaldo 455, S\~{a}o Paulo, SP, 01246-903, Brazil}


\begin{abstract}
In this paper, we analyze the temporal evolution of the age-dependent force
of infection and incidence of rubella, after the introduction of a very
specific vaccination programme in a previously nonvaccinated population
where rubella was in endemic steady state. We deduce an integral equation
for the age-dependent force of infection, which depends on a
number of parameters that can be estimated from the force of infection in
steady state prior to the vaccination program. We present the results of
our simulations, which are compared with observed data. 
We also examine the influence of contact patterns among members of
a community on the age-dependent intensity of transmission of rubella and
on the results of vaccination strategies. As an example of the
theory proposed, we calculate the effects of vaccination strategies for four
communities from Caieiras (Brazil), Huixquilucan (Mexico), Finland and the
United Kingdom. The results for each community differ considerably according
to the distinct intensity and pattern of transmission in the absence of
vaccination. We conclude that
this simple vaccination program is not very efficient (very slow) in the
goal of eradicating the disease. This gives support to a mixed strategy,
proposed by Massad \textit{et al.},
accepted and implemented by the government of the State of S\~{a}o Paulo,
Brazil.
\end{abstract}


\maketitle

\bigskip

\section{Introduction \label{sec1}}

The control of directly transmitted, viral childhood infections, around the
globe has been strongly dependent on vaccination, the most effective control
tool developed so far \cite{Plotkin99}. There are several infections for
which vaccines exist. These are therefore candidates for eradication. Some
examples include polio, measles and rubella, just to mention a few.
Vaccination strategies, however, have been each time more dependent on 
inferences based on quantitative models, which can, through simulation
tools, yield distinct scenarios and possibilities. These simulation techniques,
in turn, have been proved to be invaluable tools for helping health authorities
to decide between competitive strategies of eradication or control of those 
infections.

In previous publications \cite{Massad1994, Massad1995a}, we applied
mathematical models to design and to evaluate the impact of vaccination
against rubella in the state of S\~{a}o Paulo, Brazil. Rubella is a viral 
infection that causes a mild disease, but it is
considered to be a public health problem due to the risk of fetal
infection and subsequent congenital defects \cite{Massad1994,Azevedo94,Plotkin2001}.
Therefore, the goal of rubella vaccination is to prevent from the 
congenital rubella syndrome (CRS). 
Plotkin \cite{Plotkin2001} argues that, due to the high prevalence of
rubella in some countries, only high vaccine coverage would avoid
the increase of CRS.

In this paper, we analyse the effects of different
contact patterns on vaccination strategies against rubella
in some communities. 
We investigate a plausible form
for the contact rate function including some constrains
it must satisfy.
We concentrate on the integral equation for
the age-dependent force of infection --- defined as the 
age-dependent number of new
infections per capita, per unit time ---, which relates the pattern of contacts
among the members of a population with the prevalence of the disease,
following a methodology developed elsewhere \cite{Coutinho1993}.
The basic idea is to examine the force of infection in steady state that
results from a given vaccination strategy. 

We also turn our attention to the dynamics of the process, 
following the time development of the age-dependent force of infection 
when a vaccination
strategy is started at a certain time $t$ in a previously nonimmunized
population. Some aspects of the age and time dependences in epidemic models 
have already been studied by some authors (e.g., \cite{Greenhalgh87, Inaba90}).

This paper is organized as follows. In Sec. \ref{sec2}, 
we present the formalism
used. 
We describe in detail how the contact rate function is related to the force of
infection, discuss some constrains it must satisfy, and propose a form for it.
In Sec. \ref{sec3}, we describe the fitting procedures adopted to determine the
values of the parameters of the contact rate function for different
communities.
In Sec. \ref{sec4a}, we analyse the impact of specific vaccination
strategies against rubella using data from communities from Caieiras, a
Brazilian small town located in the neighbourhoods of S\~{a}o Paulo city
(Azevedo Neto \textit{et al. }\cite{Azevedo94}), Huixquilucan, Mexico
(Golubjatnikov \textit{et al. }\cite{Golubjatnikov71}), Finland (Edmunds 
\textit{et al. }\cite{Edmunds00}) and the United Kingdom (Farrington \textit{%
et al. }\cite{Farrington01}). It must be noted that the results
from Brazil and Mexico are from nonvaccinated communities, while the
results from Finland and the United Kingdom are from nonvaccinated males in
communitites that have partially vaccinated female population \cite{Farrington01,
Ukkonen96}. The results from S\~{a}o Paulo will be compared with those
previously reported by Massad \textit{et al.} \cite{Massad1994, Massad1995b}.
This paper differs from those quoted above, in that the form of the 
contact rate function, representing the age related pattern of contacts, was studied more
carefully. Also the relation between the vaccination rate $\nu $ and
the resulting proportion of vaccinated people [see Eqs. (\ref{55})-(\ref
{57})] was modified. In spite of this, as we shall see, the
recommended vaccination strategy was maintained, but the calculated effects
of the vaccination strategies seem now to be more realistic.
In Sec. \ref{sec4b}, we present
simulations of the temporal evolution of the force of infection and, in
Sec. \ref{sec4c}, we compare our results to experimental results. Finally, in
Sec. \ref{sec5}, we summarize our results.

\section{Mathematical Developments \label{sec2}}

\subsection{Temporal Evolution \label{subsec21}} 

Let us assume a $SIR$ model (Susceptible--Infected--Recovered). Let $%
S(a,t)da $, $I(a,t)da,$ and $R(a,t)da$ be, respectively, the number of susceptible,
infected and non-susceptibles (including recovered and vaccinated) 
individuals with ages between $a$ and $a+da$ at time $%
t$. We can write 
\begin{eqnarray}
\frac{\partial S(a,t)}{\partial a}+\frac{\partial S(a,t)}{\partial t}
&=&-[\lambda (a,t)+\nu (a,t)+\mu ]S(a,t)  \nonumber \\
\frac{\partial I(a,t)}{\partial a}+\frac{\partial I(a,t)}{\partial t}
&=&\lambda (a,t)S(a,t)-(\mu +\gamma )I(a,t)  \label{1} \\
\frac{\partial R(a,t)}{\partial a}+\frac{\partial R(a,t)}{\partial t} &=&\nu
(a,t)S(a,t)+\gamma I(a,t)-\mu R(a,t)\, ,  \nonumber
\end{eqnarray}
where $\nu (a,t)$ is the age and time-dependent rate of vaccination,  
$\gamma $ is the recovery rate, and $\mu $
is the mortality rate, assumed constant.
This type of mortality rate (constant) is known as
type-II mortality function. 
Another type of survival curve (type I) considers that 
all individuals survive to exactly a certain age,
and then die. Anderson and May \cite{Anderson91} mention 
that, for both developed and developing
countries, the observed mortality function is intermediate between
type I and type II, although closer to type I for developed regions.

The definition of the force of infection, as a function of
age and time, is 
\begin{equation}
\lambda (a,t)=\int_{0}^{\infty }da^{\prime }\beta (a,a^{\prime })\frac{%
I(a^{\prime },t)}{N(a^{\prime },t)}\quad ,  \label{2}
\end{equation}
and $N(a,t)=S(a,t)+I(a,t)+R(a,t)$ is the total number of individuals whose
ages are between $a$ and $a+da$ at time $t$.
In this equation, $\beta (a,a^{\prime })$ is the so-called contact rate
function. It is defined so that $\beta (a,a^{\prime })dada^{\prime }$ is the
number of contacts a person with age between $a$ and $a+da$ makes with all
persons with age between $a^{\prime }$ and $a^{\prime }+da^{\prime }$ per
unit time. Therefore, $\beta (a,a^{\prime })$ describes the contact patterns
among the members of a population.

Taking into account the three equations of system (\ref{1}), we can write,
for $N(a,t)$, 
\begin{equation}
\left( \frac{\partial }{\partial a}+\frac{\partial }{\partial t}\right)
N(a,t)=-\mu N(a,t)\, .  \label{5}
\end{equation}

For simplicity, we consider that, at time $t$, the total population has size 
$N$. In other words, we have taken $N(a,t)=N(a)=N(0)e^{-\mu a},$ for a given 
$t.$ In this equilibrium situation, the mortality rate equals the natality
rate, and we have $N(0)=\mu N.$

\subsubsection{Integral equation for $\protect\lambda (a,t)$ \label{subsubsec211}}

Applying the method of the characteristics, as proposed by Trucco \cite
{Trucco65} (see also \cite{Kot2000}) for solving the McKendrick-Von
Foerster equation, we can solve the system of equations (\ref{1}).

Let $s(a,t)$ and $i(a,t)$ be the proportions of susceptible and
infected individuals, among those with age $a$ at time $t$,
given by 
\begin{equation}
s(a,t)= \frac{S(a,t)}{N(a,t)} \, ,\quad i(a,t)=\frac{I(a,t)}{N(a,t)} \, .  
\label{3}
\end{equation}

With these previous definitions, the first two equations 
of the partial differential equations system (\ref{1})
can also be written as follows: 
\begin{eqnarray}
\frac{\partial s(a,t)}{\partial a}+\frac{\partial s(a,t)}{\partial t}
&=&-[\lambda (a,t)+\nu (a,t)]s(a,t)  \label{6} \\
\frac{\partial i(a,t)}{\partial a}+\frac{\partial i(a,t)}{\partial t}
&=&\lambda (a,t)s(a,t)-\gamma i(a,t)  \, .\label{7} 
\end{eqnarray}

The boundary conditions are such that, at age $a=0$, 
for $t\geq 0$, we have 
$s(0,t)=1$ and $i(0,t)=0$.
At time \mbox{$t=0$}, for $a\geq 0$,
we have that $s(a,0)$ and $i(a,0)$ are functions of age.
In the calculations, the upper limit  for the age is taken 
to be $L=60$ yr.

Considering the change of variables (as those proposed by Trucco 
\cite{Trucco65}) 
\begin{eqnarray*}
\xi  &=&a-t \\
\eta  &=&t
\end{eqnarray*}
we have 
\[
s(a,t) = s(\xi +\eta ,\eta )=s^{\prime }(\xi ,\eta ) 
\]
and similarly for $i(a,t)$, $\lambda (a,t)$ and $\nu (a,t)$.

We also have that $(\partial /\partial a+\partial /\partial t)=\partial
/\partial \eta $. Thus, taking into account the above mentioned change of
variables, Eq. (\ref{6}) reads 
\begin{equation}
\frac{\partial }{\partial \eta }\ln s^{\prime }(\xi ,\eta )=-[\lambda
^{\prime }(\xi ,\eta )+\nu ^{\prime }(\xi ,\eta )]\quad ,  \label{9}
\end{equation}
whose generic solution can be written as 
\begin{equation}
\ln s^{\prime }(\xi ,\eta )=-\int_{p}^{\eta }[\lambda ^{\prime }(\xi ,x)+\nu
^{\prime }(\xi ,x)]dx+f(\xi )\quad ,  \label{10}
\end{equation}
where $p$ and $f(\xi )$, parameters related to the boundary conditions, are
given by 
\begin{eqnarray}
f(\xi ) &=&\ln s^{\prime }(\xi ,0)\,,\quad p=0 \label{11} \\
f(\xi ) &=&\ln s^{\prime }(\xi ,-\xi )\,,\quad p=-\xi \quad .  \label{12}
\end{eqnarray}
for $\xi >0$ and $\xi <0$, respectively.

Then, for the cases in which $\xi >0$ ($a>t$) or $\xi <0$ ($a<t)$, we have,
respectively, the following solutions: 
\begin{eqnarray}
\xi >0:\ s^{\prime }(\xi ,\eta )&=&s^{\prime }(\xi ,0)\,e^{
-\int_{0}^{\eta }[\lambda ^{\prime }(\xi ,x)+\nu ^{\prime }(\xi ,x)]\,dx}
\label{13} \\
\xi <0:\ s^{\prime }(\xi ,\eta )&=&s^{\prime }(\xi ,-\xi )\,e^{
-\int_{-\xi }^{\eta }[\lambda ^{\prime }(\xi ,x)+\nu ^{\prime }(\xi ,x)]\, dx}
\ \ .  \label{14}
\end{eqnarray}

Rewriting the equations above in terms of $a$ and $t$, we obtain 
\begin{eqnarray}
s(a,t) &=&s(a-t,0)\,\exp \Bigg[-\int_{0}^{t}[\lambda (a-t+x,x)  \nonumber \\
&+&\nu(a-t+x,x)]\,dx\Bigg] \label{15} \\
s(a,t) &=&s(0,t-a)\,\exp \Bigg[-\int_{t-a}^{t}[\lambda (a-t+x,x)\nonumber \\
&+&\nu(a-t+x,x)]\,dx\Bigg]=   \nonumber \\
&=&s(0,t-a)\,\exp \Bigg[-\int_{0}^{a}[\lambda (z,z-a+t) \nonumber \\
&+&\nu (z,z-a+t)]\,dz%
\Bigg]\ \ .   \label{16}
\end{eqnarray}
for $a>t$ and $t>a$, respectively.

Equation (\ref{7}) 
\begin{equation}
\Bigg(\frac{\partial }{\partial a}+\frac{\partial }{\partial t}\Bigg)%
i(a,t)+\gamma i(a,t)=\lambda (a,t)s(a,t)  \label{17}
\end{equation}
can be rewritten, with the change of variables, as 
\begin{equation}
\frac{\partial }{\partial \eta }i^{\prime }(\xi ,\eta )+\gamma i^{\prime
}(\xi ,\eta )=\lambda ^{\prime }(\xi ,\eta )s^{\prime }(\xi ,\eta )\quad ,
\label{18}
\end{equation}
whose solution is 
\begin{equation}
i^{\prime }(\xi ,\eta )=e^{-\int_{q}^{\eta }\gamma ds}\Bigg[\int_{q}^{\eta
}dx\,\lambda ^{\prime }(\xi ,x)s^{\prime }(\xi ,x)e^{\int_{q}^{x}\gamma
ds}+g(\xi )\Bigg]\, ,  \label{19}
\end{equation}
where $q$ and $g(\xi )$ depend on the boundary conditions: 
\begin{eqnarray}
g(\xi ) &=&i^{\prime }(\xi ,0)\,,\quad q=0
\label{20} \\
g(\xi ) &=&i^{\prime }(\xi ,-\xi )\,,\quad
q=-\xi \quad .  \label{21}
\end{eqnarray}
for $\xi >0$ and $\xi <0$, respectively.

Equation (\ref{19}), in terms of $a$ and $t$, is given by 
\begin{eqnarray}
i(a,t) &=&\int_{0}^{t}dt^{\prime }\,\lambda (a-t+t^{\prime },t^{\prime
})s(a-t,0) \alpha (a,t,t^{\prime})+  \nonumber \\
&+&e^{-\gamma t}i(a-t,0),\quad a>t , \label{22} \\
i(a,t) &=&\int_{0}^{a}da^{\prime }\,\lambda (a^{\prime },a^{\prime
}-a+t)s(0,t-a) \psi (a,t,a^{\prime}) +  \nonumber \\
&+&e^{-\gamma a}i(0,t-a),\quad a<t,  \label{23}
\end{eqnarray}

\noindent where $\alpha (a,t,t^{\prime})$ and $\psi (a,t,a^{\prime})$ 
are 
\begin{equation}
\alpha (a,t,t^{\prime}) = e^{-\int_{0}^{t^{\prime }}[\lambda (a-t+\tau ,\tau )
+\nu (a-t+\tau,\tau )]d\tau }e^{\gamma (t^{\prime }-t)}\,
\end{equation}
and
\begin{equation}
\psi (a,t,a^{\prime}) = e^{-\int_{0}^{a^{\prime }}[\lambda (z,z-a+t)+\nu
(z,z-a+t)]dz}e^{\gamma (a^{\prime }-a)}\, .
\end{equation}

The age and time-dependent force of infection [Eq. (\ref{2})] can also
be written as 
\begin{equation}
\lambda (a,t)=\int_{0}^{\infty }da^{\prime }\beta (a,a^{\prime })i(a^{\prime
},t)\quad .  \label{24}
\end{equation}

In the calculations, as already explained, the upper limit of the above integral
is taken to be $L=60$ yr.
Thus, replacing solutions (\ref{22}) and (\ref{23}) for $i(a,t)$ in the above
definition, and considering that age is in the interval $0\leq a\leq L$,
the integral equation for the age and time-dependent force of infection is given by
\medskip

\begin{eqnarray}
\lambda (a,t) &=&\int_{0}^{min(t,L)}da^{\prime }\,\beta (a,a^{\prime
})\int_{0}^{a^{\prime }}da^{\prime \prime }\,\lambda (a^{\prime \prime
},a^{\prime \prime }-a^{\prime }+t) \nonumber \\
&\times &\psi(a^{\prime},t,a^{\prime \prime})
+\theta (L-t) \int_{t}^{L}da^{\prime }\,\beta (a,a^{\prime }) \nonumber \\
& \times & \Bigg[\int_{0}^{t}dt^{%
\prime }\,\lambda (a^{\prime }-t+t^{\prime },t^{\prime })s(a^{\prime }-t,0)
\alpha(a^{\prime},t,t^{\prime}) \nonumber \\ 
&+& e^{-\gamma t}\ i(a^{\prime }-t,0)\Bigg]\,\,,  \label{25}
\end{eqnarray}

\noindent where $\theta (L-t)$ is the Heaviside function. In the following 
section, we study the steady state of Eq. (\ref{25}).

\subsubsection{Steady state behavior \label{sec_bl}}

Let $S(a)da$ be the number of susceptible individuals with age between $a$
and $a+da$. The fraction of potentially infectious contacts they make
with actually infectives aged between $a^{\prime }$ 
and $a^{\prime }+da^{\prime }$ per unit time is 

\begin{equation}
S(a)da\beta (a,a^{\prime })da^{\prime }\frac{I(a^{\prime
})}{N(a^{\prime })}\quad .  \label{32}
\end{equation}

The total number of potentially infective contacts of susceptibles aged
between $a$ and $a+da$ with infectives can be obtained by integrating
Eq. (\ref{32}) in $da^{\prime }.$ Then, we obtain an expression for
the age-dependent force of infection similar to Eq. (\ref{24}).

Equation (\ref{23}) in the steady state condition gives 
\begin{eqnarray}
i(a)&=&e^{-\gamma a}\Bigg[ \int_{o}^{a} da^{\prime} e^{\gamma a^{\prime}} 
\lambda (a^{\prime}) s(0) \nonumber \\
& \times & e^{-\int_{0}^{a^{\prime}} dz 
[\lambda (z) + \nu(z)]} + i(0) \Bigg] \, \, . \label{34} 
\end{eqnarray}

Substituting this expression in the definition of 
the age-dependent force of infection in steady state we have

\begin{eqnarray}
\lambda (a)&=&\int_{0}^{\infty }da^{\prime} \beta(a,a^{\prime })
\int_{0}^{a^{\prime}}da^{\prime \prime } e^{-\gamma (a^{\prime }-a^{\prime \prime })}
\lambda (a^{\prime \prime }) \nonumber \\
& \times & e^{-\int_{0}^{a^{\prime \prime }}dz
[\lambda (z) + \nu (z)] } \quad . \label{35}
\end{eqnarray}

The integral equation (\ref{35}) always has $\lambda (a) = 0$
as solution. According to Lopez and Coutinho \cite{Lopez2000},  
depending on the parameters of the integral equation,
it may have another unique positive solution.

Equation (\ref{35}) is the limit for large $t$ of Eq. (\ref{25}):

\[
\lambda (a) = \lim_{t\to \infty }\lambda (a,t) \, .
\]

\subsection{Contact Patterns \label{sec2b}}

\subsubsection{Symmetry in the contact pattern}

As mentioned in the Introduction, one of our main difficulties is to choose a correct
form for the contact function $\beta (a,a^{\prime }).$ In this section, we
analyze a specific situation in which $\beta (a,a^{\prime })$ has to
satisfy a symmetry relation that restricts its form:
if a person $A$ has a contact with a person $B$, 
then $B$ had a contact with $A$. In terms of transmission
dynamics, it means that the total number of contacts a group $C$ of
infected individuals make with a group $D$ of susceptibles equals
the number of contacts group $D$ had with group $C$. 
This symmetry is relevant when a direct,
person-to-person contact is required for transmission. For instance, a direct
contact is required for sexually transmitted diseases. It seems to be at
least partially required for the transmission of directly transmitted
childhood diseases such as rubella. 

The number of contacts the susceptibles with age between $a$ and $a+da$ make
with infectives with age between $a^{\prime }$ and $a^{\prime
}+da^{\prime }$, in a time interval $\partial t$, is, as we have seen, 
\begin{equation}
S(a)da\beta (a,a^{\prime })da^{\prime }\frac{I(a^{\prime
})}{N(a^{\prime })}\partial t\quad .
\label{38}
\end{equation}

This number must be equal to the number of contacts the infectives with age
between $a^{\prime }$ and $a^{\prime }+da^{\prime }$ make with the
susceptibles with age between $a$ and $a+da.$ This number is 
\begin{equation}
I(a^{\prime })da^{\prime }\beta (a^{\prime },a)da\frac{S(a)}{%
N(a)}\partial t\quad .  \label{39}
\end{equation}

Thus, we must have 
\begin{equation}
S(a)\beta (a,a^{\prime })\frac{I(a^{\prime })}{N(a^{\prime })}%
=I(a^{\prime })\beta (a^{\prime },a)\frac{S(a)}{N(a)}  \label{40}
\end{equation}
or 
\begin{equation}
\frac{\beta (a,a^{\prime })}{N(a^{\prime })}=\frac{\beta (a^{\prime
},a)}{N(a)}\quad .  \label{41}
\end{equation}

Since $N(a)=N(0)e^{-\mu a}$, we see that
Eq. (\ref{41}) is satisfied if $\beta (a,a^{\prime })$ has the form 
\begin{equation}
\beta (a,a^{\prime })=e^{\mu a}h(a,a^{\prime })\quad ,  \label{42}
\end{equation}
where $h(a,a^{\prime })$ is symmetric, that is 
\begin{equation}
h(a,a^{\prime })=h(a^{\prime },a)\quad .  \label{43}
\end{equation}

Equation (\ref{42}) will be used in the following section to construct an analytical
form for $\beta (a,a^{\prime })$.

\subsubsection{A form for the contact function $\protect\beta (a,a^{\prime })$}

Let us consider that rubella is approximately transmitted by direct
person-to-person contact. In this case, considering that children are
stratified mainly by age in classrooms \cite{Massad1995a}, it is
reasonable to assume that contacts are more intense among children
with the same age. It is then convenient to write 
$h(a,a^{\prime })$ as a product of two functions,
\begin{equation}
h(a,a^{\prime })=f(a,a^{\prime })g(a,a^{\prime })\quad .  \label{44}
\end{equation}
The function $f(a,a^{\prime })$ represents the longitudinal
profile of $h(a,a^{\prime })$ along the plane $a=a^{\prime }$ and $%
g(a,a^{\prime })$ the transversal profile related to the spread of $%
h(a,a^{\prime })$ to both sides of the plane $a=a^{\prime }.$

We have chosen the following positively skewed function for $f(a,a^{\prime
}) $ 
\begin{equation}
f(a,a^{\prime })=b_{1}(a+a^{\prime })e^{-b_{2}(a+a^{\prime })}\quad ,
\label{45}
\end{equation}
and a Gaussian-like function for $g(a,a^{\prime })$ 
\begin{equation}
g(a,a^{\prime })=e^{-(a-a^{\prime })^{2}/\sigma ^{2}}\quad ,  \label{46}
\end{equation}
where $\sigma =\sigma (a,a^{\prime })$ is related to the width of the
Gaussian-like distribution to the sides of $a=a^{\prime }$. Considering a
linear spread 
\begin{equation}
\sigma (a,a^{\prime })=b_{3}+b_{4}(a+a^{\prime })\quad ,  \label{47}
\end{equation}
we obtain 
\begin{equation}
h(a,a^{\prime })=b_{1}(a+a^{\prime })e^{-b_{2}(a+a^{\prime
})}e^{-(a-a^{\prime })^{2}/[b_{3}+b_{4}(a+a^{\prime })]^{2}}\quad ,
\label{48}
\end{equation}
where $b_{1}$, $b_{2}$, $b_{3}$ e $b_{4}$ are the parameters to be
determined.

Thus, taking into account Eq. (\ref{42}), we have, for the contact
function $\beta (a,a^{\prime })$, 
\begin{equation}
\beta (a,a^{\prime })=b_{1}(a+a^{\prime })e^{-b_{2}(a+a^{\prime
})}e^{-(a-a^{\prime })^{2}/[b_{3}+b_{4}(a+a^{\prime })]^{2}}e^{\mu a}\, .
\label{49}
\end{equation}

Other functions could be chosen for $h(a,a^{\prime })$, as those proposed in
Coutinho \textit{et al. }\cite{Coutinho1993} and Massad \textit{et al. }\cite
{Massad1995a}.

\subsection{The relationship between vaccination rate and vaccine coverage}

For our next simulations, we need to define what we mean by vaccination
routine. In a nutshell, we take 
\begin{equation}
\nu (a,t)=\nu \theta (a-a_{0})\theta (a_{1}-a)\theta (t-t_{0})\quad ,
\label{50}
\end{equation}
which has the following interpretation: after time $t_{0}$ years,
children are vaccinated at a constant rate of $\nu $ children per unit of
time when their ages are between $a_{0}$ and $a_{1}$. In practice, the government
usually informs through the media that mothers should take their children to
health centers to receive the shots. The response of parents to the government
advertisement results in a given $\nu $. Enthusiastic response results in a
high $\nu $.

In steady state, Eq. (\ref{50}) becomes 
\begin{equation}
\nu (a)=\nu \theta (a-a_{0})\theta (a_{1}-a)\quad .  \label{51}
\end{equation}

We shall now calculate the relationship between $\nu $ and resulting
proportion of vaccine coverage, $p$. Let $V(a)da$ be the number of
vaccinated individuals with age between $a$ and $a+da$. Let $N_{v}(a)da$ be
the number of non-vaccinated persons with age between $a$ and $a+da$. We have 
\begin{eqnarray}
\frac{dV(a)}{da} &=&\nu (a)N_{v}(a)-\mu V(a)  \label{52} \\
\frac{dN_{v}(a)}{da} &=&-\nu (a)N_{v}(a)-\mu N_{v}(a)\quad .  \label{53}
\end{eqnarray}
Of course, we have $V(a)+N_{v}(a)=N(a)$.

Solving Eq. (\ref{52}) using the form of $\nu (a)$ given by Eq. (\ref{51}),
we have 
\begin{equation}
V(a)=\left\{ 
\begin{array}{l}
0\ \text{,} \\ 
N(0)e^{-\mu a}[1-e^{-\nu (a-a_{0})}]\ , \\ 
N(0)e^{-\mu a}[1-e^{-\nu (a_{1}-a_{0})}]\ ,
\end{array}
\begin{array}{l}
a<a_{0} \\ 
a_{0}\leq a\leq a_{1} \\ 
a>a_{1}
\end{array}
\right. \quad .  \label{54}
\end{equation}

The proportion $p$ of vaccine coverage is defined as 
\begin{equation}
p = \frac{V(a_{1})}{N(a_{1})} = 1-e^{-\nu (a_{1}-a_{0})} \quad . \label{55}
\end{equation}

The inverse relation between $\nu $ and $p$ is 
\begin{equation}
\nu =\frac{\ln (1-p)}{a_{0}-a_{1}}\quad .  \label{57}
\end{equation}

\section{Fitting the Model to The Data \label{sec3}}

Data consisted in seroprevalence studies carried out in communities from
Mexico, Brazil, Finland and the UK. 

\begin{figure}
\scalebox{0.30}{\includegraphics{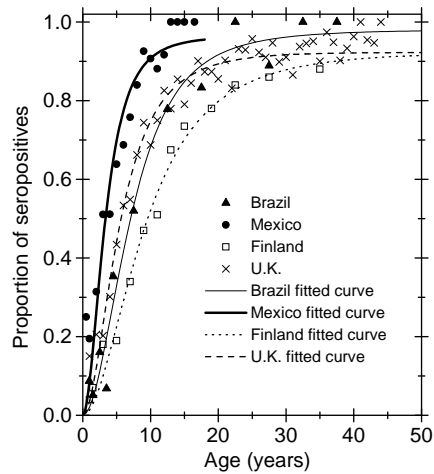}}
\caption{Seroprevalence data and corresponding fitted curves for communities from Brazil,
Mexico, Finland and the UK. The data for Caieiras (Brazil), Huixquilucan
(Mexico), Finland and the UK were taken, respectively, from the works of
Azevedo Neto \textit{et al.} \cite{Azevedo94}, Golubjatnikov \textit{et al.} 
\cite{Golubjatnikov71}, Edmunds \textit{et al. }\cite{Edmunds00} and
Farrington \textit{et al.} \cite{Farrington01}. \label{fig2}}
\end{figure}

Let $S^{+}(a)da$ be the proportion of seropositive individuals to rubella
--- whose serological tests were positive, indicating that they have already 
been infected --- with ages between $a$ and $a+da$. 
An estimate of the function $S^{+}(a)$ resulted from fitting the serological
data to (see Ref. \cite{Farrington90}) 
\begin{equation}
S^{+}(a)=1-\exp \left\{ \frac{k_{1}}{k_{2}^{2}}\left[ \left( k_{2}a+1\right)
e^{-k_{2}a}-1\right] \right\} \quad ,  \label{58}
\end{equation}
where $k_{i}$ $(i=1,2)$ are fitting parameters, estimated by the maximum
likelihood technique for all the communities except that from Finland, which
was estimated by the least squares fitting technique. Figure \ref{fig2} shows the
results of the fitting functions for the four communities considered, and
the fitting parameters are shown in Table \ref{tb:tab1}.


In our model, the seropositive individuals correspond to those who are either
infected or nonsusceptibles (recovered and vaccinated), i.e., the proportion
of seropositives, $S^{+}(a)$, is equivalent to $1-s(a)$.
The force of infection in the absence of vaccination, $\lambda _{0}(a)$, was
estimated from the seroprevalence data by the so-called 
catalytic approach (e.g., Ref. \cite{Griffiths}), according to 
\begin{equation}
\lambda _{0}(a)=\frac{dS^{+}(a)}{da}\left( 1-S^{+}(a)\right) ^{-1}\quad .
\label{59}
\end{equation}

The term catalytic arises from an analogy with chemistry. In the
dynamics of infectious diseases, an infected individual would act
as a catalyst, infecting susceptible individuals. Equation (\ref{59})
corresponds to Eq. (\ref{6}) in the steady state for the susceptible individuals,
in the absence of vaccination.

Equation (\ref{59}), expressed in terms of Eq. (\ref{58}), results in 
\begin{equation}
\lambda _{0}(a)=k_{1}a\exp \left[ -k_{2}a\right] \quad .  \label{60}
\end{equation}

The values of the parameters of the contact function $\beta (a,a^{\prime })$
[Eq. (\ref{49})] were calculated so that the resulting force of
infection $\lambda (a)$, in the absence of vaccination, obtained by solving
Eq. (\ref{35}) iteratively, agreed with $\lambda _{0}(a)$ given by
Eq. (\ref{60}). The parameters $\gamma $ and $\mu $
were taken, respectively, to be 26.0 yr$^{-1}$, corresponding to an
infectious period of 2 weeks, and 0.017yr$^{-1},$ the inverse of a life
expectancy of 60 yr. Those parameters were taken to
be the same for all communities, for simplicity. The resulting parameters of
the contact function $\beta (a,a^{\prime })$ for each community considered
are shown in Table \ref{tb:tab1}.

For Finland and the UK, we carried out simulations considering 
the two types of mortality functions described in Sec.
\ref{subsec21}. As the results were very 
similar, we discuss only those concerning type-II mortality rate.

\begin{table}[bph]
\caption{Fitting parameters ($k_{1}$ and $k_{2}$) of the seroprevalence
function, and the parameters of the contact function, for each community
considered.\label{tb:tab1}}
\begin{center}
\begin{tabular}{lcccccc}
\hline\hline
Community & $k_{1}$(yr$^{-2}$) & $k_{2}$(yr$^{-1}$) & $b_{1}$(yr$^{-2}$) & 
$b_{2}$(yr$^{-1}$) & $b_{3}$(yr) & $b_{4}$ \\ \hline
Brazil & $0.0456$ & $0.108$ & 0.658 & 0.0468 & 3.49 & 0.341 \\ 
Mexico & $0.214$ & $0.255$ & 3.54 & 0.116 & 1.04 & 0.416 \\ 
Finland & 0.0290 & 0.1068 & 0.587 & 0.0608 & 2.77 & 0.398 \\ 
UK & $0.0833$ & $0.1804$ & 1.60 & 0.0928 & 1.747 & 0.391 \\ \hline\hline
\end{tabular}
\end{center}
\end{table}


\begin{figure}[t]
\scalebox{0.30}{\includegraphics{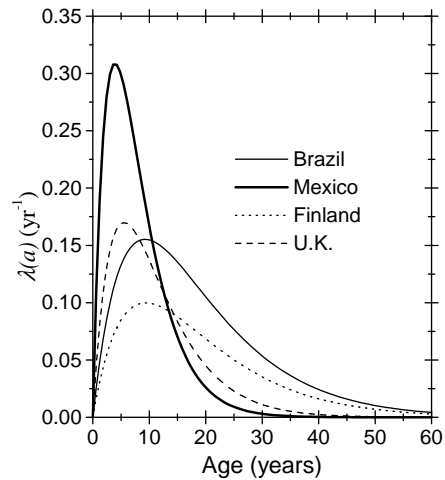}}
\caption{ Force of infection for the communities studied derived from equation
(\ref{60}).\label{fig3}}
\end{figure}

The forces of infection [as given by Eq. (\ref{60})] for
the same communities are shown in Fig. \ref{fig3}.
As can be noted, the curves have strikingly different shapes, reflecting
distinct contact patterns. As we shall see later, this has profound impact
on the calculated efficacy of different vaccination strategies.

From the force of infection, we can define the average age at which
susceptibles acquire infection 
\begin{equation}
\overline{a}=\frac{\int_{0}^{\infty }a\lambda (a)s(a)da}{\int_{0}^{\infty
}\lambda (a)s(a)da}\quad .  \label{eq:amed}
\end{equation}
We have taken the highest ages observed in the
seroepidemiological studies as the upper integration limits of the integrals
of Eq. (\ref{eq:amed}). The calculated values for the communities
studied are given in the Table \ref{tb:tab2} below.
\begin{table}[tbph]
\caption{Average age at infection for the four communities studied.
\label{tb:tab2}}
\begin{center}
\begin{tabular}{lr}
\hline\hline
Community & $\overline{a}$ (yr) \\ \hline
Caieiras, Brazil & 8.45 \\ 
Huixquilucan, Mexico & 3.96 \\ 
Finland & 10.6 \\ 
UK & 6.64 \\ \hline\hline
\end{tabular}
\end{center}
\end{table}

\begin{figure*}[t]
\scalebox{0.70}{\includegraphics{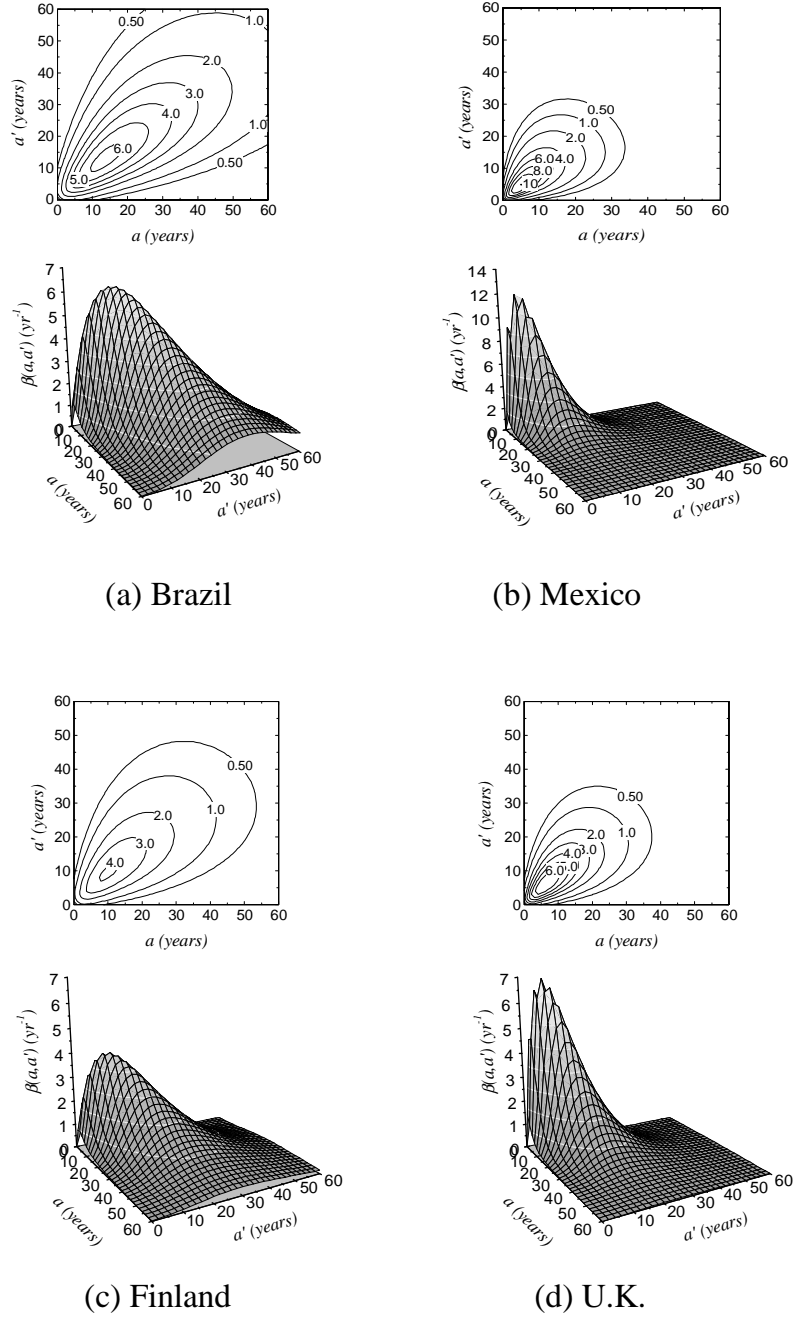}}
\caption{Calculated values of the contact
functions $\beta (a,a^{\prime })$ and respective contour plots for: (a)
Caieiras, Brazil; (b) Mexico; (c) Finland; and (d) the United Kingdom. 
\label{fig4}}
\end{figure*}

The contact functions $\beta (a,a^{\prime })$ [Eq. (\ref{49})] of the
communities considered are shown in Fig. \ref{fig4} 
as examples of the general shape
obtained. The analysis of these contact functions 
suggests two distinct patterns. In Mexico and in the United Kingdom, the age
distribution of contacts is concentrated at lower ages. In contrast, the
communities of Caieiras and Finland show a broader range of contacts, spread
over all ages. In addition, it can be noted that the density of contacts
estimated for the communities of Mexico and Caieiras are roughly twice as
high as in the United Kingdom and Finland, respectively. This may reflect
distinct social contexts between the developed and developing countries as well
as the fact that data from developed communities are only for males in
communities which have partially vaccinated female populations \cite{Ukkonen96,
Farrington01}.


\section{Simulation results \label{sec4}}

\subsection{Effects of specific vaccination strategies \label{sec4a}}

We now calculate the results of specific vaccination strategies in the above
mentioned communities, choosing $a_{0}=1$ year and $a_{1}=2$ years, $a_{0}=7$
years and $a_{1}=8$ years, and $a_{0}=14$ years and $a_{1}=15$ years for
several values of $\nu $. For a given vaccination coverage proportion $p$,
we determine $\nu$ through Eq. (\ref{57}).

The simulated results of the vaccination strategies were obtained by solving
Eq. (\ref{35}) using the values of the parameters of $\beta
(a,a^{\prime })$ obtained in Sec. \ref{sec3} for the vaccination strategies
described above.

\begin{figure}[t]
\scalebox{0.40}{\includegraphics{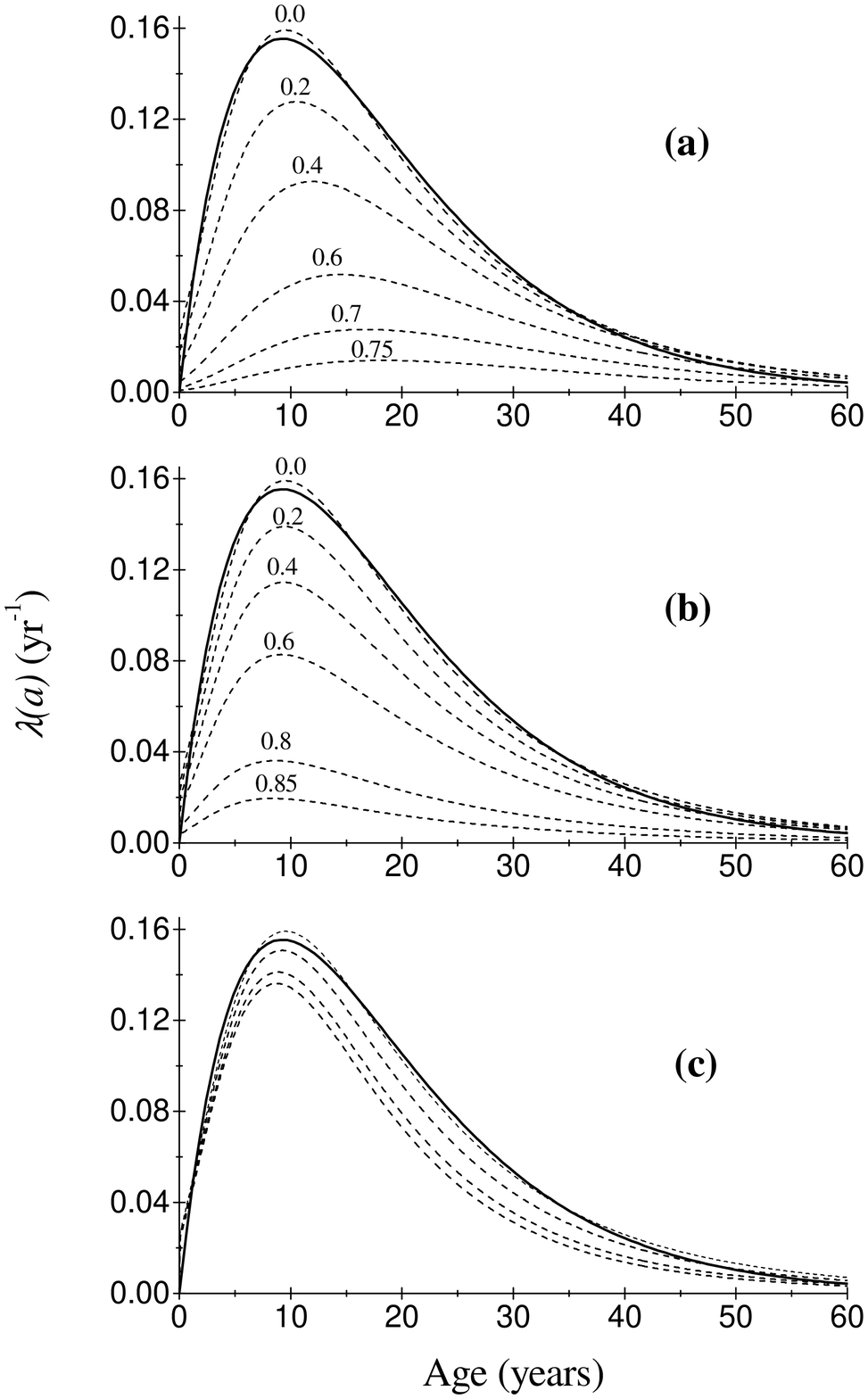}}
\caption{Effects of different vaccination programs calculated for
Caiei\-ras, Brazil. Children are vaccinated (a) between 1 and 2 yr , (b)
between 7 and 8 yr, and (c) between 14 and 15 yr. The numbers above the
dashed lines indicate the corresponding vaccine coverage and the solid lines
correspond to the catalytic model. In graph (c), the dashed lines
correspond, respectively, to 0.0, 0.4, 0.8 and 0.97 vaccine coverages.
\label{fig5}}
\end{figure}

\begin{figure}[t]
\scalebox{0.40}{\includegraphics{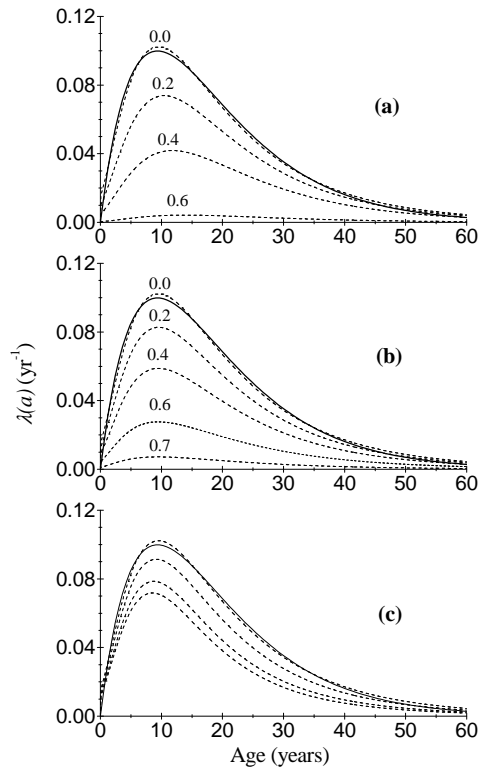}}
\caption{Effects of different vaccination programs calculated for Finland.
Children are vaccinated (a) between 1 and 2 yr, (b) between 7 and 8
yr and (c) between 14 and 15 yr. The numbers over the dashed lines
indicate the corresponding vaccine coverage and the solid lines correspond
to the catalytic model. In graph (c), the dashed lines correspond,
respectively, to 0.0, 0.4, 0.8 and 0.97 vaccine coverages.
\label{fig6}}
\end{figure}

The results for the communities of Brazil and Finland are shown,
respectively, in Figures \ref{fig5} and \ref{fig6}. The results for Mexico
and UK are not shown in graphs, but we have discussed them below.
Figure \ref{fig5} shows the results of
vaccination strategies applied to the community of S\~{a}o Paulo. Figures
4(a)--4(c) represent the different age intervals of vaccination. It
can be noted that 75\% coverage in the age interval from 1 to 2 yr almost
eliminates the disease, but the peak of infection is shifted to around 17
yr, and therefore it is displaced to the right as compared with the case
of no vaccination. A coverage between 79\% and 80\% eliminates the disease.
In Fig. 4(b), it can be noted that 85\% coverage in the age 7--8 yr
almost eliminates the disease. In addition, the peak of infection
occurs around 8 yr, and therefore it is displaced to the left as compared
with the case of no vaccination. A 90\% coverage in this age interval
eliminates the disease. Finally, Figure 4(c) shows that vaccination in the
interval from 14 to 15 yr is almost useless, since 97\% coverage has very
little impact in the force of infection, and it is impossible to eliminate
the disease, even if a 100\% coverage is used. 


Figure \ref{fig6} shows the results of vaccination strategies applied to the
community in Finland. Figures 5(a)--5(c) represent the different
age intervals of vaccination. It can be noted that 60\% coverage in the age
interval 1--2 yr [Fig. 5(a)] almost eliminates the disease, and
the age of the peak of infection is not affected at all. 
A coverage of 64\% eliminates
the disease. In Fig. 5(b), it can be noted that 70\% coverage in the age
interval 7--8 yr almost eliminates the disease, and again does not
shift the age of the peak in the force of infection. Finally, Fig. 5(c)
shows that vaccination in the interval from 14 to 15 yr is almost
useless, since 97\% coverage has very little impact on the force of
infection, and it is impossible to eliminate the disease even if a 100\%
coverage is used. 


For the Huixquilucan community in Mexico a 74\%
coverage in the age interval 1--2 yr eliminates
the disease. 
Even a 97\% coverage in the age interval
7--8 yr is not able to eliminate the disease, and indeed causes very
little effect on its force of infection.

For the community in the UK, a 66\% coverage in the age
interval 1--2 yr eliminates the disease.
Even a 97\% coverage in the 
age interval 7--8 yr is
not able to eliminate the disease. However, the peak of the force of infection
curve shifts leftwards to around 5 yr.

As expected, the results of vaccinating in the interval from 7 to 8 yr of
age are disappointing if compared to the results of vaccinating from 1 to 2
yr of age, and vaccinating between 14 and 15 yr is almost useless.

Vaccination programmes against rubella were implemented in many
countries (e.g., Refs. \cite{Edmunds00, Ukkonen96, Massad1994, Massad1995b, Odette,
MMWR}). However, vaccination coverages and strategies 
sometimes changed from one period to another.
As mentioned by Ukkonen \cite{Ukkonen96}, UK (1970) and
Finland (1975) chose selective vaccination of 11- and 13-year-old girls to 
prevent rubella and such a strategy was not effective in eradicating the virus.
These observed results agree with our simulation for vaccination from 14 
to 15 yr of age. In 1998, rubella vaccine was introduced in Mexico into the 
childhood vaccination schedule at age 1 and 6 yr \cite{MMWR}, resulting in an
intense decrease in the rubella incidence, in agreement with our simulations
for vaccination from 1 to 2 yr of age.

\subsection{Temporal evolution \label{sec4b}}
The simulations for the temporal evolution of the force of
infection were based on the numerical solutions of the
integral equation for $\lambda (a,t)$, using the parameters
of $\beta (a,a^{\prime})$ for the Caieiras community.

\begin{figure}
\scalebox{0.40}{\includegraphics{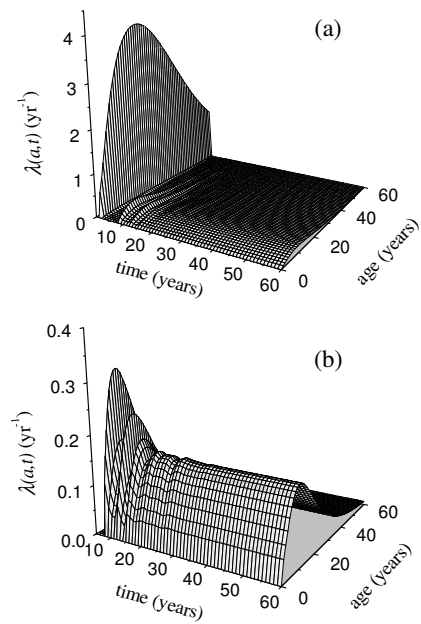}}
\caption{(a) Simulations for $\lambda (a,t)$, without vaccination,
considering a completely susceptible population, at time $t=0$, except for a
proportion 10$^{-5}$ of individuals with ages between 40 and 45 yr that
suddenly become infected; (b) the same as (a), but with the time scale
starting at $t=5$ yr, so that the initial peak is cut and the resulting
steady state is observable.\label{fig7}}
\end{figure}

Our first simulation considered a completely susceptible population (which
is not the case with Caieiras). We then assumed that, at time $t=0$, a
proportion $p_{i}=10^{-5}$ of individuals with ages between 40 and 45 yr
suddenly become infected. The resulting dynamics of the disease is shown in
Fig. \ref{fig7}.


\begin{figure}[h]
\scalebox{0.30}{\includegraphics{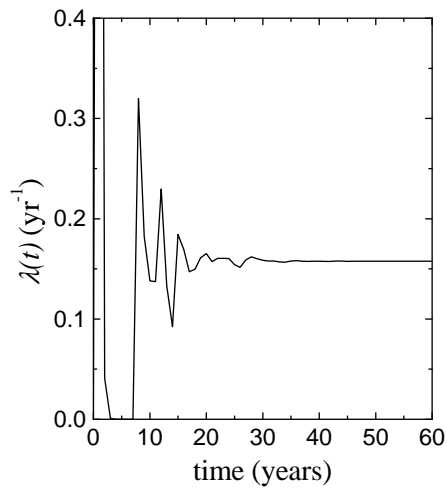}}
\caption{Profile of Fig. \ref{fig7}, cut at 8 yr age. The initial outbreak peak
between $t=0$ and $t=3$ yr almost exhausts the susceptible fraction of the
population. It takes about 3.5 years for the number of new susceptibles to
accumulate in sufficient number to trigger a second outbreak which
eventually stabilizes at an endemic steady state.\label{fig8}}
\end{figure}

It can be noted that after a few oscillations the force of infection tends
to the function $\lambda _{0}(a)$. We can also see that from around 40 yr
onwards the force of infection stabilizes. Figure \ref{fig8} displays a profile cut
at 8 yr old.


\begin{figure}
\scalebox{0.30}{\includegraphics{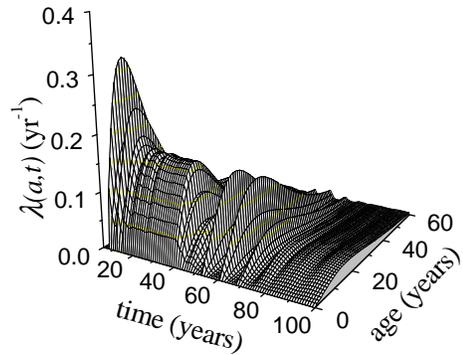}}
\caption{Simulations for $\lambda (a,t),$ with the same initial conditions
considered in Fig. \ref{fig7}, but including a vaccination routine of form (\ref
{50}), with $a_{0}=1$ yr, $a_{1}=2$ yr, $t_{0}=40$ yr, and a
vaccination coverage of 70\%.\label{fig9}}
\end{figure}

For our next simulation, the same conditions as the above simulation were
applied, and a vaccination routine of form (\ref{50}) with $a_{0}=1$
yr, $a_{1}=2$ yr, $t_{0}=40$ yr, and a vaccination coverage of 70\%
was added. The results for all ages are 
shown in Fig. \ref{fig9}. It can be noted that after the
introduction of the vaccination, the force of infection oscillates before
reaching a steady state, much lower than $\lambda_{0}(a)$. 
The whole process takes around 40 yr to reach the new
steady state. The fact that the
process takes so long to reach a steady state does not recommend this
vaccination strategy for eradicating the disease.

\begin{figure}[b]
\scalebox{0.30}{\includegraphics{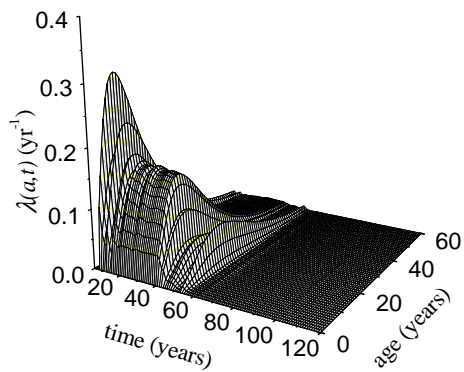}}
\caption{Vaccination scheme considered in the simulations of Fig. 
\ref{fig9}, but with a vaccination coverage of 80\%.\label{fig11}}
\end{figure}

The next simulation uses the same vaccination scheme, but with a vaccination
coverage of 80\%. The results for all ages are shown in Fig. \ref{fig11}.

For this coverage one can see that the disease is eradicated in
approximately 20 yr. Again the fact that the process takes so long does
not recommend this vaccination strategy for control.

\subsection{Comparison of specific features with real data \label{sec4c}}

\begin{figure}
\scalebox{0.35}{\includegraphics{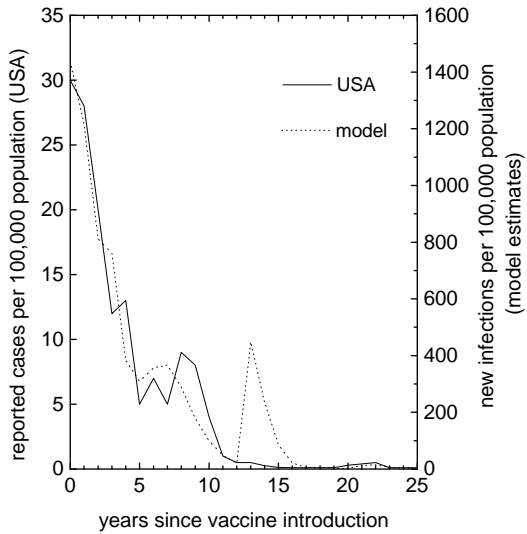}}
\caption{Impact on the number of reported cases (left-hand scale) of the
vaccination strategy against rubella adopted in the U.S. in 1969
(data taken from CDC \cite{CDC94}), together with the results of our
simulation (right-hand scale) for a vaccination rate that results in a 80\%
vaccination coverage.\label{fig13}}
\end{figure}

The strategy given by Eq. (\ref{50}) was actually adopted in the United
States in 1969 \cite{Plotkin99}. The results of the impact on the incidence
is shown in Fig. \ref{fig13} (left-hand scale) together with the results of our
simulation (right-hand scale) for a $\nu $ that results in a 80\% coverage.
The model estimates for the number of new infections per 100\,000 population
were calculated according to the following equation:
\begin{equation}
Y(t)=\frac{1}{N}\int_{0}^{\infty }da\ \lambda (a,t)S(a,t)\quad ,  \label{30}
\end{equation}

\noindent where $\int_{0}^{\infty }da\ \lambda (a,t)S(a,t)$ is the number of new
cases per unit time at time $t$. In the calculations, the upper limit of the 
above integral was taken to be 60 yr.

It should be noted that the incidence calculated from the seroprevalence
data is two orders of magnitude larger than the incidence that results from
notification. In fact, it is known that only a fraction of all infections
display the clinical features of rubella disease. In addition, only a
fraction of those rubella cases is officially notified. However, several
qualitative features of the data are quite similar to those observed in
the simulations described in the preceding section. Let us comment, in more
detail, on the more significant similarities. 

In 1977, that is, 8 yr after the introduction of the program, it was
noted that although the program was having a major impact on rubella
in children, rubella rates in those older than 15 yr were not
substantially different from prevaccination rates. We shall see now that
this effect is shown in our simulations.

\begin{figure}
\scalebox{0.35}{\includegraphics{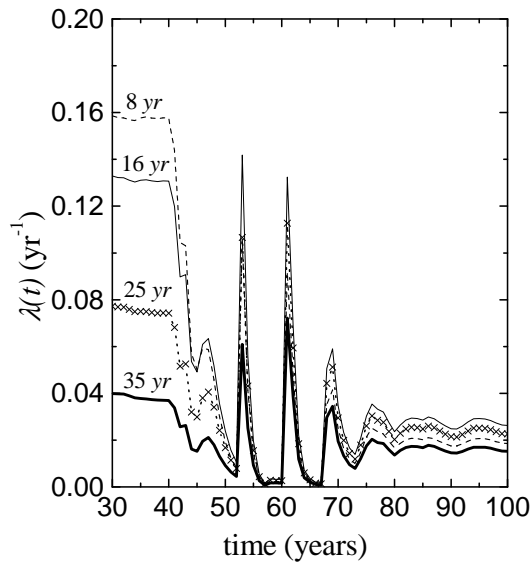}}
\caption{Profile of Fig. \ref{fig9}, cut at the ages of 8, 16, 
25 and 35 yr.\label{fig14}}
\end{figure}

Figure \ref{fig14} represents four cuts of Fig. \ref{fig9}, corresponding 
to 70\% coverage, at the
ages of 8, 16, 25 and 35 yr. It can be seen that the above mentioned
effect is clearly observed. The drop in the force of infection at the age of
8 yr is much larger than at 16 and 25 yr, and the effect at the age of
35 yr is almost negligible. 

However, about 15 years after the introduction of the vaccine, three major
outbreaks are observed in the simulations, and this may be dangerous. 
The pattern of several oscillations in the incidence of an infectious
disease, after the introduction of vaccination, has already been 
observed in real data \cite{rub}. If the
vaccinal coverage is increased to 80\%, after small outbreaks, the disease
disappears, as shown in Fig. \ref{fig11}. 

\section{Summary \label{sec5}}

In this paper, we analyzed the temporal evolution of the age-dependent force
of infection and incidence of rubella, after the introduction of a very
specific vaccination program in a previously nonvaccinated population
where rubella was in an endemic steady state. This very specific vaccination
program consists in vaccinating children within a certain age range with a
rate determined essentially by the public response to government advertisements.

We conclude that a simple vaccination program is not very efficient
(very slow) in the goal of eradicating the disease. This gives support to a
mixed strategy proposed by Massad \textit{et al.} \cite{Massad1994},
accepted and implemented by the government of the State of S\~{a}o Paulo.
This strategy recommended a mass vaccination campaign against rubella in the
State of S\~{a}o Paulo for all children with ages between 1 and 10 yr as
an initial intervention followed by a vaccination program of the form
given by (\ref{50}), in the routine calendar at 15 months of age. As reported
in Refs. \cite{Massad1995b,Plotkin2001}, the
results were very good, and there was a considerable reduction in the number 
of rubella and congenital rubella syndrome cases. The incidence of rubella 
and CRS remained at low levels with the routine vaccination program, 
in agreement with our simulation results for high vaccination coverages.

We have also applied a formalism developed elsewhere 
\cite{Coutinho1993} to calculate the effects of vaccination routines designed to
reduce or eliminate rubella.

This formalism provides an integral equation for the force of infection in a
steady state given the pattern of contacts between the members of the
population and the specific form of the vaccination routines.

To apply the formalism, the pattern of contacts between the members of the
population, the so-called contact function $\beta (a,a^{\prime })$, has
to be estimated. Some symmetries obeyed by $\beta (a,a^{\prime })$ and a
general form for it were studied in Sec. \ref{sec2b}. In Sec. \ref{sec3}, 
the force of
infection in the absence of the vaccination was calculated from
seroprevalence data from four communities in Brazil, Mexico, Finland and the
United Kingdom. With this force of infection, in the absence of vaccination,
the contact function $\beta (a,a^{\prime })$ for each community was
estimated. It was noted that $\beta (a,a^{\prime })$ differed considerably
between the communities studied, which is in agreement with the differences in the
force of infection in the absence of vaccination.

Finally, in Sec. \ref{sec4a}, the effects of several vaccination routines were
calculated for the four communities studied. As a general conclusion, one
can say that vaccination between 1 and 2 yr presents distinct advantages
over any other strategy considered. In Caieiras, vaccination between 7 and
8 yr has the apparent advantage of shifting the average age of the first
infection leftwards. However, if the coverage is above 60\%, the impact of
vaccinating between 1 and 2 yr on the force of infection is twice as high
as vaccinating between 7 and 8 yr. This result confirms our previous
analysis and recommendations of 1992 (Massad \textit{et al. }\cite
{Massad1994}). In all other communities studied, vaccination between 7 and 8
yr results in very disappointing impact when compared with vaccination
between 1 and 2 yr.

\begin{acknowledgments}
We would like to thank the anonymous referee for his/her very helpful
suggestions and comments. We acknowledge support from FAPESP and PRONEX/CNPq.
\end{acknowledgments}


\begin{thebibliography}{99}
\bibitem{Plotkin99}  S. A. Plotkin and W. A. Orenstein (editors), \textit{%
Vaccines} (Saunders, USA, 1999).

\bibitem{Massad1994}  E. Massad, M. N. Burattini, R. S. Azevedo Neto, H. M.
Yang, F. A. B. Coutinho, and D. M. T. Zanetta, Epidemiol. Infect. \textbf{112%
}, 579 (1994).

\bibitem{Massad1995a}  E. Massad, R. S. Azevedo Neto, H. M. Yang, M. N.
Burattini, and F. A. B. Coutinho, J. Biol. Syst. \textbf{3}, 803 (1995).

\bibitem{Azevedo94}  R. S. Azevedo Neto, A. S. B. Silveira, D. J. Nokes , H.
M. Yang, S. D. Passos, M. R. A. Cardoso, and E. Massad, Epidemiol. Infect. 
\textbf{113}, 161 (1994).

\bibitem{Plotkin2001} S. A. Plotkin, Vaccine \textbf{19}, 3311 (2001).

\bibitem{Coutinho1993}  F. A. B. Coutinho, E. Massad, M. N. Burattini, H. M.
Yang, H. M., and R. S. Azevedo Neto, IMA J. Math. Appl. Med. Biol. \textbf{10%
}, 187 (1993).

\bibitem{Greenhalgh87}  D. Greenhalgh, IMA J. Math. Appl. Med. Biol. \textbf{%
4}, 109 (1987).

\bibitem{Inaba90}  H. Inaba, J. Math. Biol. \textbf{28}, 411 (1990).

\bibitem{Golubjatnikov71}  R. Golubjatnikov, W. R. Elsea, and L. Leppla, Am.
J. Trop. Med. Hyg. \textbf{20}, 958 (1971).

\bibitem{Edmunds00}  W. J. Edmunds, N. J. Gay, M. Kretzschmar, R. G. Pebody,
and H. Wachmann, Epidemiol. Infect. \textbf{125}, 635 (2000).

\bibitem{Farrington01}  C. P. Farrington, M. N. Kanaan, and N. J. Gay, J.
Roy. Stat. Soc. C (Appl. Stat.) \textbf{50}, 251 (2001).

\bibitem{Ukkonen96}  P. Ukkonen, Scand. J. Infect. Dis. \textbf{28}, 31
(1996).

\bibitem{Massad1995b}  E. Massad, R. S. Azevedo-Neto, M. N. Burattini, D. M.
T. Zanetta, F. A. B. Coutinho, H. M. Yang, J. C. Moraes, C. S. Panutti, V.
A. U. F. Souza, A. S. B. Silveira, C. J. Struchiner, G. W. Oselka, M. C. C.
Camargo, T. M. Omoto, and S. D. Passos, Int. J. Epidemiol. \textbf{24 }(4),
842 (1995).

\bibitem{Anderson91}  R. M. Anderson and R. M. May, \textit{Infectious
Diseases of Humans: Dynamics and Control} (Oxford University Press, Oxford,
1991).

\bibitem{Trucco65}  E. Trucco, Bull. Math. Biophys. \textbf{27}, 285 (1965).

\bibitem{Kot2000}  M. Kot, \textit{Elements of Mathematical Ecology}
(Cambridge University Press, Cambridge, 2000).

\bibitem{Lopez2000} L. F. Lopez and F. A. B. Coutinho, J. Math. Biol. 
\textbf{40}, 199 (2000).

\bibitem{Farrington90}  C. P. Farrington, Stat. Med.\emph{\ }\textbf{9}, 953
(1990).

\bibitem{Griffiths} D. A. Griffiths, Appl. Statist. \textbf{23}, 330 (1974).

\bibitem{Odette} O. G. van der Heijden, M. A. E. Conyn-van Spaendock,
A. D. Plantinga, and M. E. E. Kretzschmar, Epidemiol. Infect. \textbf{121},
653 (1998).

\bibitem{MMWR} Centers for Disease Control and Prevention (CDC),
MMWR \textbf{49 }(46): 1048 (2000). 

\bibitem{CDC94}  Centers for Disease Control and Prevention (CDC), 
MMWR \textbf{43 }(21): 397 (1994).

\bibitem{rub} E. Massad, R. S. Azevedo, M. N. Burattini,
D. M. T. Zanetta, and F. A. B. Coutinho (unpublished).

\end{thebibliography}
\end{document}